\documentclass[twocolumn,prl,superscriptaddress,amsfonts]{revtex4-1}
\usepackage{physics}
\usepackage{graphicx}

\renewcommand{\figurename}{Figure}
\makeatletter
\renewcommand*{\fnum@figure}{{\normalfont\bfseries \figurename~\thefigure}}
\renewcommand*{\@caption@fignum@sep}{ $|~$}
\usepackage[pdftex,bookmarks=true,bookmarksopen,bookmarksnumbered,
                colorlinks,
                linkcolor=blue,
                citecolor=blue]{hyperref}
\usepackage{float}
\usepackage{newfloat}

\DeclareFloatingEnvironment[
    fileext=los,
    name=Figure,
    placement=tbhp,
    within=none,
]{extfig}
\renewcommand{\extfigname}{Extended Data Figure}
\renewcommand*{\fnum@extfig}{{\normalfont\bfseries \extfigname~\theextfig}}

\begin{document} 
  

\title{Silicon quantum processor unit cell operation above one Kelvin}

\author{C. H. Yang}
\email[]{henry.yang@unsw.edu.au}
\author{R. C. C. Leon}
\author{J. C. C. Hwang}
\altaffiliation[Now at ]{Research and Prototype Foundry, The University of Sydney, NSW 2006, Australia.}
\author{A. Saraiva}
\author{T. Tanttu}
\author{W.~Huang}
\affiliation{
 Centre for Quantum Computation and Communication Technology,
 School of Electrical Engineering and Telecommunications,
 The University of New South Wales, Sydney, NSW 2052, Australia
}
\author{J. Camirand Lemyre}
\affiliation{
 Institut Quantique et D\'epartement de Physique, 
 Universit\'e de Sherbrooke, Sherbrooke, Qu\'ebec J1K 2R1, Canada
}
\author{K. W. Chan}
\affiliation{
 Centre for Quantum Computation and Communication Technology,
 School of Electrical Engineering and Telecommunications,
 The University of New South Wales, Sydney, NSW 2052, Australia
}
\author{K. Y. Tan}
\affiliation{
QCD Labs, QTF Centre of Excellence, 
Department of Applied Physics, 
Aalto University, 00076 Aalto, Finland
}
\author{F. E. Hudson}
\affiliation{
 Centre for Quantum Computation and Communication Technology,
 School of Electrical Engineering and Telecommunications,
 The University of New South Wales, Sydney, NSW 2052, Australia
}
\author{K. M. Itoh}
\affiliation{
 School of Fundamental Science and Technology,
 Keio University, 3-14-1 Hiyoshi, Kohokuku, Yokohama 223-8522, Japan.
}
\author{A. Morello}
\affiliation{
 Centre for Quantum Computation and Communication Technology,
 School of Electrical Engineering and Telecommunications,
 The University of New South Wales, Sydney, NSW 2052, Australia
}
\author{M. Pioro-Ladri\`ere}
\affiliation{
 Institut Quantique et D\'epartement de Physique,
 Universit\'e de Sherbrooke, Sherbrooke, Qu\'ebec J1K 2R1, Canada
}
\affiliation{
 Quantum Information Science Program,
 Canadian Institute for Advanced Research,
 Toronto, ON, M5G 1Z8, Canada
}
\author{A. Laucht}
\affiliation{
 Centre for Quantum Computation and Communication Technology,
 School of Electrical Engineering and Telecommunications,
 The University of New South Wales, Sydney, NSW 2052, Australia
}
\author{A. S. Dzurak}
\email[]{a.dzurak@unsw.edu.au}
\affiliation{
 Centre for Quantum Computation and Communication Technology,
 School of Electrical Engineering and Telecommunications,
 The University of New South Wales, Sydney, NSW 2052, Australia
}

\maketitle
 


\textbf{
Quantum computers are expected to outperform conventional computers for a range of important problems, from molecular simulation to search algorithms, once they can be scaled up to large numbers of quantum bits (qubits), typically millions~\cite{Feynman1982,Loss1998,Fowler2012}. For most solid-state qubit technologies, e.g. those using superconducting circuits or semiconductor spins, scaling poses a significant challenge as every additional qubit increases the heat generated, while the cooling power of dilution refrigerators is severely limited at their operating temperature below 100~mK~\cite{Devoret2004,Vandersypen2017,Almudever2017}.
Here we demonstrate operation of a scalable silicon quantum processor unit cell, comprising two qubits confined to quantum dots (QDs) at $\boldsymbol{\sim}$1.5~Kelvin.
We achieve this by isolating the QDs from the electron reservoir, initialising and reading the qubits solely via tunnelling of electrons between the two QDs~\cite{Bertrand2015,Veldhorst2017,Crippa2018}. We coherently control the qubits using electrically-driven spin resonance (EDSR)~\cite{Pioro2008,Leon2019} in isotopically enriched silicon $\boldsymbol{^{28}}$Si~\cite{itoh_watanabe_2014}, attaining single-qubit gate fidelities of 98.6\% and coherence time $\boldsymbol{T_2^*}$ = 2~$\boldsymbol{\mu}$s during `hot' operation, comparable to those of spin qubits in natural silicon at millikelvin temperatures ~\cite{Takeda2016,Kawakami2016,Watson2018,Zajac2018}. Furthermore, we show that the unit cell can be operated at magnetic fields as low as 0.1~T, corresponding to a qubit control frequency of 3.5~GHz, where the qubit energy is well below the thermal energy.
The unit cell constitutes the core building block of a full-scale silicon quantum computer, and satisfies layout constraints required by error correction architectures~\cite{Veldhorst2017,Jones2018}.
Our work indicates that a spin-based quantum computer could be operated at elevated temperatures in a simple pumped $\boldsymbol{^4}$He system, offering orders of magnitude higher cooling power than dilution refrigerators, potentially enabling classical control electronics to be integrated with the qubit array~\cite{Hornibrook2015,Degenhardt2017}.
}

\begin{figure*} 
	\centering 
		\includegraphics[page=1,trim=1.4cm 18cm 1.4cm 0cm]{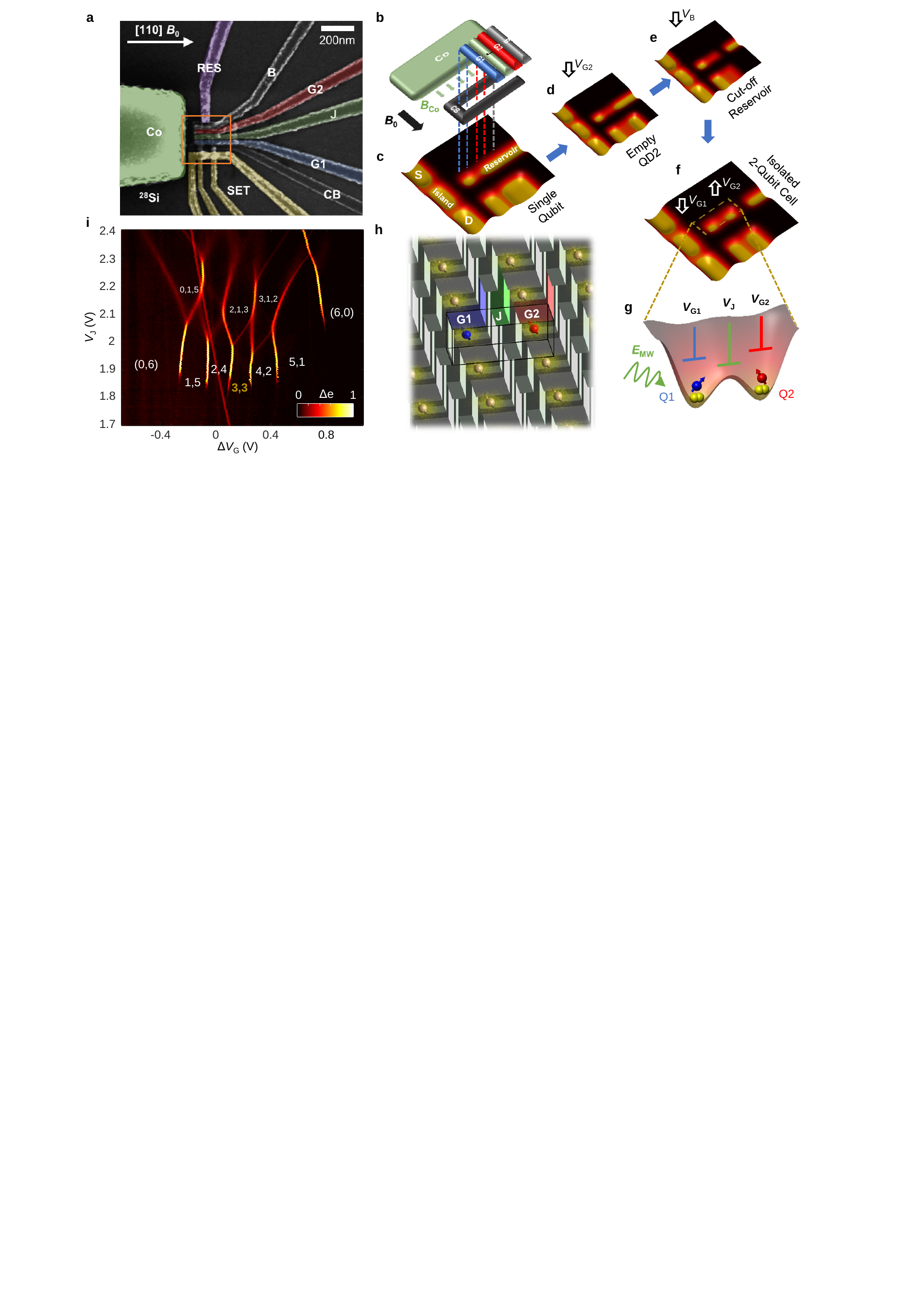}
		\caption{\textbf{An isolated spin qubit processor unit cell.}
		\textbf{a}, SEM image of an identical two-qubit device with Co micromagnet for EDSR control~\cite{Leon2019}.
		\textbf{b}, Schematic of the Al gate stack. QDs are defined under G1 and G2, and laterally confined by CB. J controls the coupling between the QDs, while B can be biased to create a barrier between the QDs and the electron reservoir. The Co micromagnet provides a magnetic field gradient while simultaneously delivering a microwave voltage signal to enable EDSR. Charge sensor (SET) and electron reservoir (RES) are not shown in this schematic.  
		\textbf{c-f}, Tuning sequence to obtain an isolated (3,3) electron configuration.
		\textbf{c}, Starting with a single dot under G1 as in Ref.~\cite{Leon2019}, we load 6 electrons from the reservoir onto QD1.
		\textbf{d}, We lower $V_\text{G2}$ just enough to deplete all electrons under G2.
		\textbf{e}, We lower $V_\text{B}$ from 3.2~V to 0~V to create a barrier that makes it almost impossible for electrons to escape.
		\textbf{f}, We re-bias $V_\text{G1}$ and $V_\text{G2}$ to define QD2 under G2, and move 3 electrons from QD1 to QD2.
		\textbf{g}, Schematic of the conduction band and control electrodes of the isolated qubit unit cell in the (3,3) charge configuration. Electron spins in excited valley states are used for qubit operation.
		\textbf{h}, Schematic of a qubit unit cell within a large-scale 2D quantum processor. The unit cell occupies the minimum foot print for operating a two-qubit system. Scaling towards pairwise unit cell operation allows construction of a complete quantum computer. 
		\textbf{i}, Charge stability diagram of the isolated QDs, with a total of 6 electrons trapped in the system. Here, $\Delta V_\text{G}= V_\text{G1}-V_\text{G2}$ and $V_\text{G1}+V_\text{G2}=4.8$~V. The system evolves into a strongly coupled three-dot system for very positive biasing of $V_\text{J}$. The charge transition near $\Delta V_\text{G}=0$~V is not coupled to the QDs, and most likely corresponds to charge movement outside the CB confinement area.
		Tilting the double QD potential at low $V_{\text J}<2$~V allows us to set any charge configuration between (0,6) and (6,0), while a high $V_{\text J}>2$~V transforms the double QD system into a triple QD system, with a third dot forming under gate J.
		}
		\label{fig:device}
\end{figure*}

Electrostatically gated QDs in Si/SiGe or Si/SiO$_2$ heterostructures are prime candidates for spin-based quantum computing due to their long coherence times, high control fidelities, and industrial manufacturability~\cite{Veldhorst2014,Kawakami2016,Marurand2016,Takeda2016,Yoneda2018,Yang2019}. 
In large scale quantum processors the qubits will be arranged in either 1D chains~\cite{Jones2018} or 2D arrays~\cite{Fowler2012} to enable quantum error correction schemes. For architectures relying on exchange coupling for two-qubit operation~\cite{Veldhorst2015,Watson2018,Zajac2018,Huang2018},
the QDs are expected to be densely packed. 
Until now, two-qubit QD systems have been tunnel-coupled to a nearby charge reservoir that has typically been used for initialisation and readout using spin-to-charge conversion~\cite{Elzerman2004}. 
Here we demonstrate an isolated double QD system that requires no tunnel-coupled reservoir~\cite{Bertrand2015,Veldhorst2017,Crippa2018} to perform full two-qubit initialisation, control and readout -- thus realising the elementary unit cell of a scalable quantum processor (see \autoref{fig:device}h). 

\autoref{fig:device}a shows a scanning electron microscope (SEM) image of a silicon metal-oxide-semiconductor (Si-MOS) double QD device nominally identical to the one measured. The device is designed with a cobalt micromagnet to facilitate EDSR, whereby an AC voltage at frequency $f_\text{qubit}$ is applied to the micromagnet electrode to drive spin resonance~\cite{Pioro2008}, and a single electron transistor (SET) charge sensor is used to detect changes in the electron occupation of the two QDs~\cite{Leon2019}. The experimental setup is described in \autoref{fig:meas_setup}.
In \autoref{fig:device}b-f we illustrate the tuning sequence that we use to configure the isolated double QD unit cell in the (3,3) charge configuration.
We start by accumulating the desired total number of electrons under G1, then deplete the electrons under gate J and G2, and finally cut off the electron reservoir by lowering the bias applied to the barrier gate B.
At the end of the tuning sequence, the strong barrier confinement ensures no electrons can tunnel into or out of the qubit cell. 
The ability to operate the unit cell without any changes in electron occupation throughout initialization, control and readout is the prerequisite for scaling it up to large 2D arrays (see \autoref{fig:device}h), where qubit control can be achieved by global magnetic resonance or via an array of micromagnets that allow local EDSR.
Using the gates G1, J and G2 (see \autoref{fig:device}g) we can distribute the 6 electrons arbitrarily within the qubit cell, as demonstrated in the stability diagram shown in \autoref{fig:device}i. In this work, we focus on the (3,3) charge configuration (See \autoref{fig:3e3e}).
Here, the lower two electrons in each dot form a spin-zero closed shell in the lower conduction-band valley state, and we use the spins of the unpaired electrons in the upper valley states of the silicon QDs as our qubits~\cite{Veldhorst2015b}. It is also possible to operate the qubits in the (1,1) and (1,3) charge configurations (see \autoref{fig:1e3e}), but (3,3) is chosen for better EDSR driving strength and J-gate control~\cite{Leon2019}.

\begin{figure*} 
	\centering 
		\includegraphics[page=2,trim=1.4cm 14cm 1.4cm 0cm]{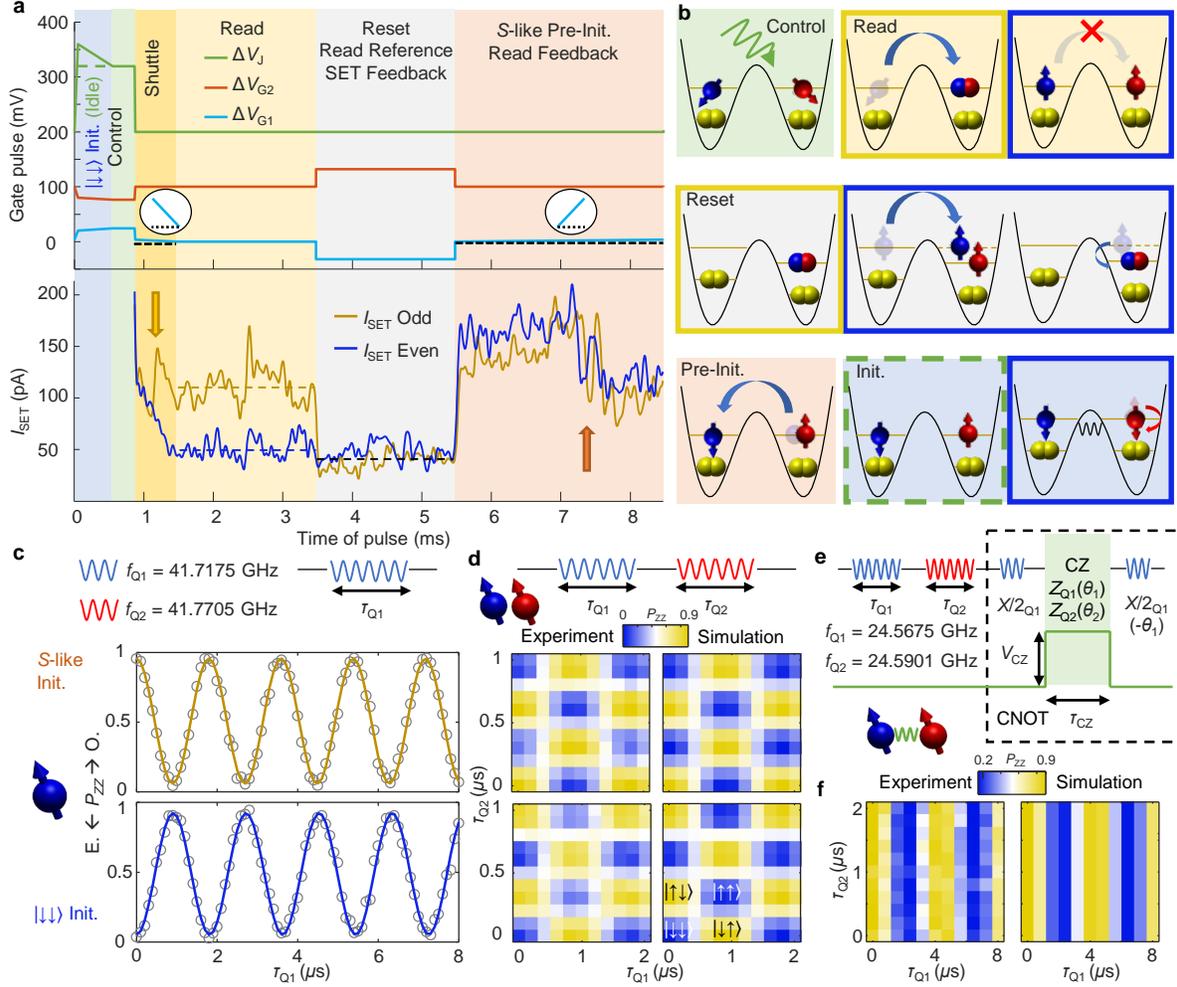}
		\caption{\textbf{Full two-qubit operation in an isolated quantum processor unit cell.}
		\textbf{a},\textbf{b} Operational sequence for two-qubit control, readout, reset, calibration, and initialisation using single-shot parity readout without any nearby reservoir. Gate pulses and SET current are displayed in \textbf{a}, while the corresponding charge movement is depicted in \textbf{b}.
		After the qubit control stage, parity readout, based on Pauli spin blockade when trying to shuttle the blue electron to Q2, is conducted. If blockade occurs, no charge movement event happens and the SET exhibits a lower current (blue $I_\text{SET}$), and a higher current (yellow $I_\text{SET}$) when the electron is not blockaded.
		The reset stage forces the blue electron to move to Q2 regardless of its spin state, by pulsing deep enough to overcome the Pauli spin blockade. The system subsequently relaxes into the (2,4) $S$ state. $I_\text{SET}$ measured here is used as reference for the parity readout. Next, the blue electron is moved back to Q1 via a slow ramp, ensuring a $S$-like state is initialised.
		Finally, the system can be left in the $S$-like state (odd parity spin) or initialised in the $T_-$ state ($\left |\downarrow\downarrow \right \rangle$) if pulsed to a hot-spot relaxation region prior to control. 
		\textbf{c}, Rabi oscillations of Q1 with initialisation in the $S$-like (top panel) and $\left |\downarrow\downarrow \right \rangle$ (bottom panel) states. During readout, even parity is mapped to low signal (=0) and odd parity to high signal (=1). The fitted Rabi oscillation amplitude is $\sim90$~\%. Measured at 40~mK and 1.4~T.
        \textbf{d}, Serially driven Rabi rotations on two independent (uncoupled) qubits and subsequent parity readout for $S$-like (top panel) and $\left |\downarrow\downarrow \right \rangle$ (bottom panel) initialisation. For $\left |\downarrow\downarrow \right \rangle$ initialisation, the oscillations span the separable two-qubit space with $\left |\downarrow\downarrow \right \rangle$ and $\left |\uparrow\uparrow \right \rangle$ giving low signal, and $\left |\downarrow\uparrow \right \rangle$ and $\left |\uparrow\downarrow \right \rangle$ giving high signal. 
        The simulations (right panels) further validate that the parity readout does indeed follow the expectation value of the $\hat{P}_{ZZ}=\frac{1}{2}(\hat{\sigma}_{II}-\hat{\sigma}_{ZZ})$ projection. Measured at 40~mK and 1.4~T.
        \textbf{e}, Sequence to implement a CNOT two-qubit logic gate via a controlled-Z (CZ) rotation using J-gate control. The final $\pi/2$-pulse incorporates a phase shift to compensate the Stark shift from pulsing the J-gate.
        \textbf{f}, Applying a CNOT gate before readout turns the parity readout into single qubit spin readout. Starting with $\left |\downarrow\downarrow \right \rangle$ initialisation, when the CNOT gate is applied after the serial two-qubit Rabi oscillations from \textbf{d}, we measure only the Rabi oscillations of Q1. Tiny oscillations (period = 1~$\mu$s) in the experimental data (left panel) along the $y$-direction can still be observed due to imperfect CZ pulsing. Simulations are shown in the right panel. Measured at 40~mK and 0.8~T. 
		}
		\label{fig:parity}
\end{figure*}

\begin{figure*}[t]
	\centering 
		\includegraphics[page=3,trim=1.4cm 19cm 1.4cm 0cm]{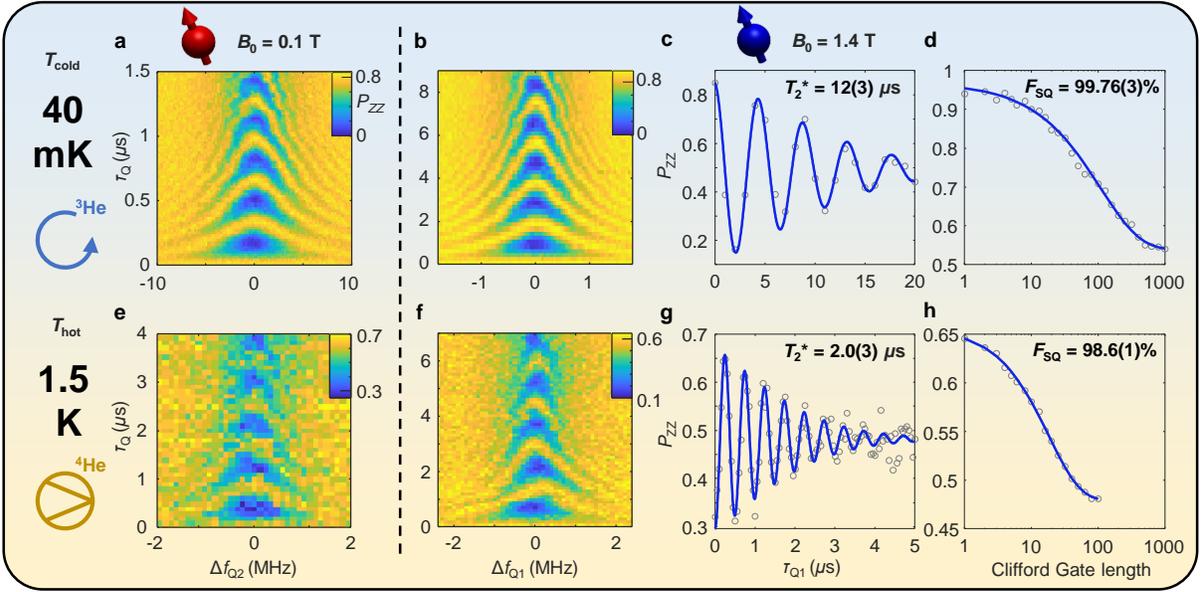}
		\caption{\textbf{Qubit operation at high temperature and low magnetic field.}
		\textbf{a-d}, $^3$He circulation on, with mixing chamber temperature at $T_\text{cold} = 40$~mK.
		\textbf{a}, Rabi chevron of Q2 at $B_0 = 0.1$~T, where $f_\text{Q2}$ = 3.529~GHz.
		\textbf{b}, Rabi chevron of Q1 at $B_0 = 1.4$~T, where $f_\text{Q1}$ = 41.71~GHz.
		\textbf{c}, Ramsey coherence time, $T_2^*$ at $B_0 = 1.4$~T.
		\textbf{d}, Randomised benchmarking performance at $B_0 = 1.4$~T.
		\textbf{e-h}, Pumped $^4$He only, with $T_\text{hot} = 1.5$~K.
		\textbf{e}, Rabi chevron of Q2 at $B_0 = 0.1$~T, where $f_\text{Q2}$ = 3.535~GHz, and $hf_\text{Q2} \ll k_\text{B}T$ = 125~$\mu$eV.
		\textbf{f}, Rabi chevron of Q1 at $B_0 = 1.4$~T, where $f_\text{Q1}$ = 41.71~GHz.
		\textbf{g}, Ramsey coherence time, $T_2^*$ at $B_0 = 1.4$~T.
		\textbf{h}, Randomised benchmarking performance at $B_0 = 1.4$~T. 
		Error range of the benchmark numbers are within 95\% confidence level.
		See Methods section for details of Ramsey measurements.
		The randomised benchmarking protocol is identical to the one used in Ref. \cite{Veldhorst2014}.
		}
		\label{fig:hot}
\end{figure*}

We depict the entire control, measurement and initialisation cycle in \autoref{fig:parity}a,b. Throughout operation, the same six electrons stay within the unit cell. We measure the two-spin state based on a variation of the Pauli spin blockade. As for traditional singlet-triplet readout~\cite{Ono2002}, tunnelling of the electrons into the same dot is only allowed for a spin singlet state due to the Pauli exclusion principle. On the other hand, not all triplets are blockaded -- the $T_0$ triplet mixes with the singlet state at a rate faster than our SET charge readout. Therefore any combination of $\left |\uparrow\downarrow\right \rangle$ and $\left |\downarrow\uparrow\right \rangle$ will be allowed to tunnel. As a result, spin-to-charge conversion in our device manifests itself as spin parity readout, measuring the $\hat{\sigma}_{ZZ}$ projection of the two-qubit system, where $\hat{\sigma}$ is the Pauli operator (see Methods).

In the remainder of the paper, we denote this parity readout output as $P_{ZZ}$, the expectation value of $\hat{P}_{ZZ}=\frac{1}{2}(\hat{\sigma}_{II}-\hat{\sigma}_{ZZ})$. An even spin state readout then leads to $P_{ZZ} = 0$ and an odd state leads to $P_{ZZ} = 1$. 

Initialisation is based on first preparing the unit cell in the (2,4) $S$ state, before moving one electron to Qubit 1 (Q1) to create a (3,3) $S$-like state. For $hf_\text{qubit} \gg k_\text{B} T$ we can also initialise the system in the well-defined $\left |\downarrow\downarrow \right \rangle$ state by dwelling at a spin relaxation hot-spot~\cite{Watson2018,Huang2018}. In \autoref{fig:parity}c we show Rabi oscillations for the two different initialisation states, starting in either the $S$-like or the $\left |\downarrow\downarrow \right \rangle$ state. 
Additional verification of the initialised states is performed by spin relaxation measurements described in Methods section and \autoref{fig:t1}.

We confirm that our readout procedure distinguishes the state parity by serially driven Rabi rotations shown in \autoref{fig:parity}d, where we coherently and unconditionally rotate first Q1 and then Q2, and measure the output state. 
Reading other two-qubit projections is also straightforward. 
Rotating one of the qubits by $\pi/2$ we gain access to $\hat{P}_{ij}=\frac{1}{2}(\hat{\sigma}_{II}-\hat{\sigma}_{ij})$, where $i,j \in X,Y,Z$. 
Furthermore, adding a conditional two-qubit gate like a CNOT before readout, we can turn the parity readout into a single qubit readout. \autoref{fig:parity}e shows the pulse sequence from \autoref{fig:parity}d with an added CNOT gate based on performing a conditional-Z (CZ) gate. 
We achieve the conditional phase shift by pulsing the J gate to temporarily increase the $J$ coupling between the two qubits (without changing the charge detuning). The single qubit readout result is shown in \autoref{fig:parity}f. 
The sequence reads out only the Q1 spin state as $\hat{P}_{Z1}=\frac{1}{2}(\hat{\sigma}_{II}-\hat{\sigma}_{ZI})$, independent of Q2. 
To read out Q2, one would simply need to swap target and control of the CNOT gate. 
The small but visible oscillations along the $y$-axis in the data are due to imperfect CZ pulsing. 
Details of the CNOT gate data are shown in \autoref{fig:cnot}, where the CNOT gate parameters in panel c are the same as those for \autoref{fig:parity}f. 

\begin{figure}[t!] 
	\centering 
		\includegraphics[page=4,trim=6.1cm 17cm 6.1cm 0cm]{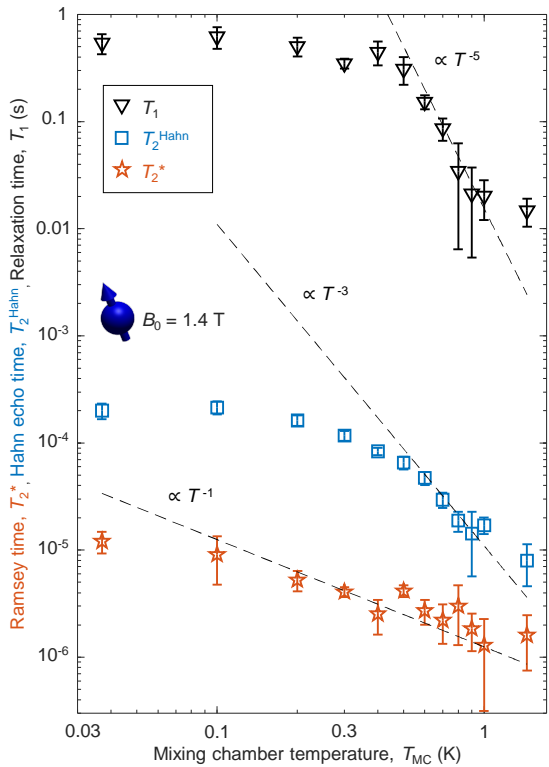}
		\caption{\textbf{Temperature dependence of qubit properties.}
		Spin relaxation time $T_1$, Hahn echo coherence time $T_2^\text{Hahn}$, and  Ramsey coherence time $T_2^*$ of Q1 as a function of mixing chamber temperature at $B_0=1.4$~T. The dotted lines indicate trends of $\propto T^{-5}$ for $T_1$, $\propto T^{-3}$ for $T_2^\text{Hahn}$, and $\propto T^{-1}$ for $T_2^*$.
		Error bars represent the 95\% confidence level.
		}
		\label{fig:t1t2temp}
\end{figure}

Having demonstrated the general operation of the quantum processor unit cell including initialisation, one- and two-qubit control, and parity and single-qubit readout, we can now investigate the effect of temperature. For large scale quantum computer integration, the benefits of raising the temperature to reduce engineering constraints have to be carefully balanced with the presence of increased noise. Prior studies have examined the relaxation of Si-MOS QD spin qubits at temperatures of 1.1~K~\cite{Petit2018} and coherence times of ensembles of Si-MOS QDs up to 10~K~\cite{Shankar2010}. The coherence times of single deep-level impurities in silicon at 10~K~\cite{Ono2019} and ensembles of donor electron spins in silicon up to 20~K~\cite{Tyryshkin2003,Tyryshkin2012} were also examined.
However, coherence times and gate fidelities of these qubits have not been investigated as yet. 
Here we investigate the gate fidelity of a fully-controllable spin qubit at 1.5~Kelvin.

In \autoref{fig:hot} we present single-qubit Rabi chevrons and randomised benchmarking for temperatures of $T_{\text{cold}}=40~\pm~5$~mK in \autoref{fig:hot}a-d, and $T_{\text{hot}}=1.5~\pm~0.1$~K in \autoref{fig:hot}e-h. Here, $T_{\text{hot}}=1.5$~K is achieved by simply pumping on the $^4$He in the 1K pot of the dilution refrigerator, while the $^3$He circulation is completely shut off. Qubit operation and readout at this elevated temperature is possible since our QDs have relatively high valley splitting ($>500$~$\mu$eV) and orbital splitting energies ($> 2.5$~meV)~\cite{Leon2019}. We observe Rabi chevrons, indicating coherent qubit control, for both $B_0 = 1.4$~T and $B_0 = 0.1$~T at $T_{\text{hot}}=1.5$~K, despite the thermal energy being larger than the qubit energy ($k_\text{B}T \gg hf_{\text{qubit}}$).

From the decay of the Ramsey oscillations in \autoref{fig:hot}g we determine a coherence time $T_2^*$ at this elevated temperature of $2.0\pm0.3$~$\mu$s, comparable to that in natural silicon at mK temperatures~\cite{Takeda2016,Kawakami2016,Watson2018,Zajac2018}. The single qubit gate fidelity extracted from randomised benchmarking is $F_{\text{SQ}}=98.6\pm0.1$~\%, nearly at the fault tolerant level (see \autoref{fig:hot}h). For reference, the qubit's performance at $T_{\text{cold}}=40$~mK is shown in \autoref{fig:hot}a-d, at which both $T_2^*$ and $F_{\text{SQ}}$ are about 6 times better.

We present a more detailed study of the coherence times and relaxation times as a function of mixing chamber temperature $T_\text{MC}$ in \autoref{fig:t1t2temp}.  A similar study as a function of external magnetic field is presented in \autoref{fig:t1t2}, where we observe the Hahn echo time to scale linearly with $B_0$, where shorter relaxation times at lower field are possibly due to spin-orbit Johnson noise~\cite{struck2019spin}. Temperature has the strongest impact on $T_1$, which scales as $T^{-5}$ between 0.5 and 1.0 K. This could be interpreted as a Raman process involving intervalley piezophonons stemming from the oxide layer -- if the spin-lattice relaxation was dominated by Si deformation potential phonons, the temperature power law should be stronger, as discussed in Ref.~\cite{Petit2018}. $T_2^\text{Hahn}$ and $T_2^*$ display a weaker dependence on temperature. 

The results in \autoref{fig:t1t2temp} show a significant reduction in spin relaxation and coherence times going from 100~mK to 1.5~K. While this reduction does not prevent the qubits from being operated at this temperature, future device engineering should aim to minimise possible sources of noise for optimised high-temperature operation. Residual $^{29}$Si nuclear spins that couple to the qubits through the hyperfine interaction lead to background magnetic field noise that could be easily reduced by using silicon substrates with higher isotopic enrichment~\cite{Witzel2010}. Our devices contain $800$~ppm residual $^{29}$Si atoms, which is more than one order of magnitude higher than what is currently available~\cite{Tyryshkin2012,itoh_watanabe_2014}. 
While the gradient magnetic field from a micromagnet, as required for EDSR operation~\cite{Pioro2008,Kawakami2016,Leon2019}, might freeze out nuclear spin dynamics~\cite{Tyryshkin2012}, it will also make the qubits more sensitive to electric field noise induced by the artificial spin-orbit coupling~\cite{struck2019spin}. Charge noise has been shown to increase with temperature~\cite{Petit2018}, and could constitute the dominant noise source for EDSR systems at elevated temperatures. 
Furthermore, since the drop in visibility in \autoref{fig:hot} can be attributed to a lower charge readout fidelity owing to the broadening of the SET peak (see \autoref{fig:histogram}), replacing SET current readout with a readout mechanism that offers better signal to noise ratios should improve readout fidelities at higher temperatures. Radio-frequency gate dispersive readout~\cite{West2019,Crippa2018} could act as a solution, while, at the same time, providing a truly scalable unit cell footprint.

In conclusion, we have presented a fully operable two-qubit system in an isolated quantum processor unit cell, which allows operation up to 1.5~K -- a temperature that is conveniently achieved using pumped $^4$He cryostats -- where we reach near fault-tolerant single-qubit fidelities.
These results pave the way for scaling of silicon-based quantum processors to very large numbers of qubits.


\bibliographystyle{naturemag}
\bibliography{mybib}

\section*{Methods}\label{sec_methods}

\subsection*{Feedback controls}
Three types of feedback/calibration processes are implemented for the experiments:
\begin{itemize}
    \item{\textit{SET sensor current feedback} -- For each current trace acquired by the digitizer, $I_\text{SET}$ during the Reset stage is compared against a set value. In case of \autoref{fig:parity}a this set value is 50~pA. The SET top-gate voltage is then adjusted to ensure $I_\text{SET}$ stays at $\sim$50~pA.}
    \item{\textit{Charge detuning feedback} -- The charge detuning level between the two dots is controlled by monitoring the bias at which the charge transition occurs, shown by the red arrow in \autoref{fig:parity}a. Adjusting the bias on $V_\text{G1}$, the charge transition is then retuned to occur at 60\% of the Read Feedback stage.}
    \item{\textit{Spin qubit resonance calibration} -- Frequency calibration of the microwave frequencies is  applied during the measurement of \autoref{fig:hot}d,h. The calibration protocol is the same as the one detailed in Ref. \cite{Huang2018}.}
\end{itemize}

\subsection*{Temperature control}
For operation at base temperature $T_\text{MC} = 40~\text{mK}$, the circulation of $^3$He is fully enabled. For $T_\text{MC} > 40~\text{mK}$ and  $T_\text{MC} <= 1~\text{K}$, we turn on the heater at the mixing chamber stage (See \autoref{fig:meas_setup}) with a Proportional Integral (PI) computer controller. For $T_\text{MC} = 1.5~\text{K}$, the $^3$He circulation is completely stopped by closing the circulation valves, and turning off all heaters. The fridge is then left for at least 1 day for $T_\text{MC}$ to saturate at 1.5~K, the temperature of the 1K pot stage. The 1K pot was actively pumped during all the measurements in this work.

To validate the temperature accuracy, we performed effective electron temperature measurements of the isolated QDs by measuring the broadening of the (2,4)-(3,3) charge transition as shown in \autoref{fig:temperature}a. Having determined the lever arms from magnetospectroscopy (see \autoref{fig:magneto}) we fit the charge transitions to extract the effective electron temperatures (see \autoref{fig:temperature}b). For $T_\text{MC}>0.4$~K, the extracted temperature matches well with the mixing chamber thermometer of the fridge.

\subsection*{Wait-time-dependent phase Ramsey measurement}
To extract $T_2^*$ times at high temperatures, where the control pulses are of similar duration as the coherence time, the conventional way of setting a resonance frequency detuning would greatly suppress the already low visibility of the oscillations. A more efficient way to extract $T_2^*$ is to use zero-detuning pulses while applying a large wait-time-dependent phase shift to the second $\pi/2$-pulse. This results in fast Ramsey fringes while maintaining maximum visibility. For example, in \autoref{fig:hot}g, the phase of the second microwave pulse has the dependency
$\theta_\text{MW} = 2\times 10^6\times 2\pi\tau_\text{wait}$,
giving Ramsey fringes with a frequency of 2~MHz.

For all Ramsey measurements, each data point consist of 100 single shots per acquisition, with 5 overall repeats, giving a total of 500 single shots.

\subsection*{Parity readout}
In general, the joint state of a pair of spins may be measured through a spin-to-charge conversion based on the Pauli exclusion principle. In a double dot system, interdot tunnelling is stimulated by detuning the energy levels of one quantum dot with respect to the other. If the pair of electrons is in the singlet state, tunnelling will occur and the charge distribution in the double dot will change. A charge measurement then allows us to distinguish a singlet state from any one of the spin triplets.

For simplicity, we will refer to the possible charge configurations as $(1,1)\rightarrow(0,2)$, but any configuration with two effective valence spins is valid, including the (3,3) charge configuration  investigated in this work. If Pauli spin blockade occurred in the standard way, we would have the simple mapping
\begin{eqnarray}
    |{(1,1)S}\rangle&\rightarrow&|{(0,2)S}\rangle,\\
    |{(1,1)T_0}\rangle&\rightarrow&|{(1,1)T_0}\rangle,\\
    |{(1,1)T_+}\rangle&\rightarrow&|{(1,1)T_+}\rangle,\\
    |{(1,1)T_-}\rangle&\rightarrow&|{(1,1)T_-}\rangle,
\end{eqnarray}
and the final state after the measurement would be a pure state. A measurement of the charge state would distinguish singlets from triplets, \textit{i.e.}, discern between distinct eigenstates of total angular momentum $\hat{S}_{\rm TOT}^2=(\hat{S_1}+\hat{S_2})^2$.

In practice, relaxation between the triplet states and the singlet ground state occurs at the same time as the charge measurement process. Spin flip relaxation is  slower than 10~ms for all temperatures studied here, as shown in \autoref{fig:t1t2temp} of the main text, so that the $T_+$ and $T_-$ states will be preserved for a sufficiently long time for our SET-based measurement to be completed.
    
On the other hand, the $T_0$ triplet and the singlet are constantly mixing with each other, either through the difference in Overhauser fields from nuclear spins (reduced here in isotopically enriched $^{28}$Si), difference in g-factors under external applied magnetic field, or due to the micromagnet field gradient.
    
The wavefunction component of $|(1,1)T_0\rangle$ that mixes into the $|(1,1)S\rangle$ state rapidly relaxes into the $|(0,2)S\rangle$ state. This means that at the time scale of the $S-T_0$ mixing, the population in $|(1,1)T_0\rangle$ is depleted and relaxes into $|(0,2)S\rangle$.
    
Since this relaxation mechanism is much faster than any spin flip mechanism, after a sufficiently long time (compared to the $S-T_0$ mixing rate and the tunnel/charge relaxation rate), the $T_0$ state has completely relaxed into the singlet state and the mapping connects again two pure states
\begin{eqnarray}
    |{(1,1)S}\rangle&\rightarrow&|{(0,2)S}\rangle,\\
    |{(1,1)T_0}\rangle&\rightarrow&|{(0,2)S}\rangle,\\
    |{(1,1)T_+}\rangle&\rightarrow&|{(1,1)T_+}\rangle,\\
    |{(1,1)T_-}\rangle&\rightarrow&|{(1,1)T_-}\rangle.
\end{eqnarray}

Now, a charge measurement can distinguish between states of parallel spin state or anti-parallel spins, represented by the observable $\hat{P}_{ZZ}=\frac{1}{2}(\hat{\sigma}_{II}-\hat{\sigma}_{ZZ})$. This measurement therefore corresponds to a parity read-out.

\subsection{Confirming initialisations using spin relaxation}
By preparing a $|{\downarrow\downarrow}\rangle$ state we measure the spin relaxation time $T_1$ for both qubits by selectively flipping one of them to a spin up state, followed by a wait time, $\tau_{\rm wait}$. \autoref{fig:t1}a-c are fitted using a simple decay equation $A\exp(-\tau_{\rm wait}/T_1)+C$, when driving a, Q1 to spin up, b, off-resonance drive, and c, Q2 to spin up. The two qubits have $T_1$ = 540~ms and 36~ms, respectively.

When we repeat the same measurement with a $S$-like initialisation, we observe a mixed decay pattern. We now need to fit to a more complicated equation that measures the parity of the spins while both spins are relaxing, assuming no knowledge of the initial state.

For $|{\uparrow\downarrow}\rangle$ and $|{\downarrow\uparrow}\rangle$ components, the relaxation equations for parity readout are:
    \begin{eqnarray}\label{eq:T1-1}
    P_{ZZ|{\uparrow\downarrow}\rangle}=A_{|{\uparrow\downarrow}\rangle}e_1+C,\\
    P_{ZZ|{\downarrow\uparrow}\rangle}=A_{|{\downarrow\uparrow}\rangle}e_2+C,
    \end{eqnarray}  
where
    \begin{eqnarray}
    e_1 = \exp(-\tau_{\rm wait}/T_\text{1Q1}),\\  
    e_2 = \exp(-\tau_{\rm wait}/T_\text{1Q2}).
    \end{eqnarray}  
For a $|{\uparrow\uparrow}\rangle$ component, assuming no two-spin interactions, it is then:
\begin{equation}
    P_{ZZ|{\uparrow\uparrow}\rangle}= -2A_{|{\uparrow\uparrow}\rangle}(e_1-\frac{1}{2})(e_2-\frac{1}{2})+\frac{1}{2}+C.
\end{equation}
Combining the three equations above, for an arbitrary initial state fitting we obtain:
\begin{equation}
\begin{split}
    P_{ZZ} = &~(A_{|\uparrow\uparrow\rangle}+A_{|\uparrow\downarrow\rangle})e_1+(A_{|\uparrow\uparrow\rangle}+A_{|\downarrow\uparrow\rangle})e_2\\
    & -2A_{|\uparrow\uparrow\rangle}e_1e_2+C.\\
\end{split}
\label{eq:T1}
\end{equation}

\autoref{eq:T1} is then applied to fit \autoref{fig:t1}d-f, which then gives the probability of each eigenstate for $S$-like initialisation, proving indeed it is an equal mixture of $|{\uparrow\downarrow}\rangle$ and $|{\downarrow\uparrow}\rangle$ states.

\section*{Acknowledgements}
We acknowledge support from the US Army Research Office (W911NF-17-1-0198), the Australian Research Council (CE170100012), Silicon Quantum Computing Proprietary Limited, and the NSW Node of the Australian National Fabrication Facility. The views and conclusions contained in this document are those of the authors and should not be interpreted as representing the official policies, either expressed or implied, of the Army Research Office or the U.S. Government. The U.S. Government is authorised to reproduce and distribute reprints for Government purposes notwithstanding any copyright notation herein. K.~M.~I. acknowledges support from Grant-in-Aid for Scientific Research by MEXT. J.~C.~L. and M.~P.-L. acknowledge support from the Canada First Research Excellence Fund and in part by the National Science Engineering Research Council of Canada. K.~Y.~T. acknowledges support from the Academy of Finland through project Nos. 308161, 314302 and 316551.

\section*{Author contributions}
C.H.Y designed and performed the experiments.
C.H.Y., R.C.C.L. and A.S analysed the data. 
J.C.C.H. and F.E.H. fabricated the device with A.S.D's supervision.
J.C.C.H., T.T. and W.H contributed to the preparation of experiments.
J.C.L., R.C.C.L., J.C.C.H., C.H.Y. and M.P.-L. designed the device.
K.W.C., K.Y.T. contributed to discussions on the nanofabrication process.
K.M.I. prepared and supplied the $^{28}$Si epilayer.
T.T., W.H., A.M. and A.L. contributed to results discussion and interpretation.
C.H.Y., A.S., A.L. and A.S.D. wrote the manuscript with input from all co-authors.

\begin{extfig*} 
	\centering 
		\includegraphics[page=5,trim=1.4cm 10.5cm 1.4cm 0cm]{figures_arxiv.pdf}
		\caption{\textbf{Experimental setup.}
        The device measured is identical to the one described in Ref.~\cite{Leon2019}. It is fabricated on an isotopically enriched 900~nm thick $^{28}$Si epilayer~\cite{itoh_watanabe_2014}  with 800 ppm residual concentration of $^{29}$Si using multi-layer gate-stack silicon MOS technology~\cite{Angus2007,Lim2009}.
        Stanford Research System (SRS) SIM928 rechargeable isolated voltage source modules (within SRS SIM900 mainframes) are used to supply all our DC voltages, and a LeCroy ArbStudio 1104 arbitrary waveform generator (AWG) is combined with the DC voltages through resistive voltage dividers, with 1/5 division for DC and 1/25 for AWG inputs. The resistance of the voltage dividers in combination with the capacitance of the coaxial cables limits the AWG bandwidth to $\sim5$~MHz. Filter boxes with lowpass filtering (100~Hz for DC lines and 80~MHz for fast lines) and thermalisation are mounted at the mixing chamber plate. Shaped microwave pulses are delivered by an Agilent E8267D vector signal generator, employing its own internal AWG for in-phase/quadrature (IQ) modulation. There are two DC blocks and two attenuators along the microwave line as indicated in the schematic. The SET sensor current signal is amplified by a FEMTO transimpedance amplifier DLPCA-200 and a SRS SIM910 JFET isolation amplifier with gain of 100, before passing a SRS SIM965 lowpass filter, and finally being acquired by an Alazar ATS9440 digitizer. The SpinCore PBESR-PRO-500 acts as the master trigger source for all other instruments.
        The device sits inside an Oxford Kelvinox 100 wet dilution refrigerator with base temperature of $T_\text{MC}  = 40$~mK. The superconducting magnet is powered by an American Magnet Inc. AMI430 power supply. 
		}
		\label{fig:meas_setup}
\end{extfig*}

\begin{extfig*} 
	\centering 
		\includegraphics[page=6,trim=1.4cm 20.5cm 1.4cm 0cm]{figures_arxiv.pdf}
		\caption{\textbf{Qubit spectra in the (3,3) charge configuration region.}
		EDSR spectra of Q1 (lower frequency) and Q2 (upper frequency) as a function of $\Delta V_\text{G}= V_\text{G1}-V_\text{G2}$, measured using adiabatic microwave pulses with a sweep range of 2~MHz. The bending of Q2's spectrum suggests strong mixing with an excited state. Near the (4,2) region, both spectra split up equally due to the increase of $J$ coupling. A small splitting can also been seen near the (2,4) region. At the (2,4) and (4,2) electron charge transitions, we no longer have a proper effective two spin system and the signal vanishes. We operate our qubits mostly near the (2,4) side (left) for faster EDSR control over Q1.
		}
		\label{fig:3e3e}
\end{extfig*}

\begin{extfig*} 
	\centering 
		\includegraphics[page=7,trim=1.4cm 13.5cm 1.4cm 0cm]{figures_arxiv.pdf}
		\caption{\textbf{Qubit spectra in other charge configurations.}
		\textbf{a}, \textbf{b}, EDSR spectra of the \textbf{a}, (3,3) and \textbf{b}, (1,3) charge configurations as a function of $\Delta V_\text{G}= V_\text{G1}-V_\text{G2}$ at $B_0$ = 0.5~T.
		Between (3,3) and (1,3), the number of electrons in Q1 changes but remains constant in Q2. While the bending spectrum exhibits minimal change in frequency and can be attributed to Q2, the straight spectrum has shifted by more than 50~MHz, confirming that it corresponds to Q1. The large change in frequency of Q1 is mainly due to the unpaired electron spin now occupying the other valley state.
		\textbf{c}, For large $V_\text{J}$ a third QD starts forming under the J-gate (compare \autoref{fig:device}), and the device can be operated as two-qubit system with two electrons in the (1,0,1) and (0,1,1) configurations at $B_0$ = 1.4~T. Only one qubit resonance is clearly found, while the other one is only weakly observed when $J$ coupling increases (red circle), where spin-orbit coupling is stronger for the tightly confined dot. Insert: J-coupling increases with $V_\text{J}$, demonstrating J control when moving one electron from the (1,0,1) to the (0,1,1) charge configuration.
		}
		\label{fig:1e3e}
\end{extfig*}

\begin{extfig*} 
	\centering 
		\includegraphics[page=8,trim=1.4cm 19cm 1.4cm 0cm]{figures_arxiv.pdf}
		\caption{\textbf{Spin relaxation measurements using parity readout.}
		\textbf{a-c}, $P_{ZZ}$ with $\left |\downarrow\downarrow \right \rangle$ initialisation and \textbf{a}, flipping spin of Q1 adiabatically ($\left |\uparrow\downarrow \right \rangle$), \textbf{b}, no spin flip ($\left |\downarrow\downarrow \right \rangle$), and \textbf{c}, flipping spin of Q2 adiabatically ($\left |\downarrow\uparrow \right \rangle$).
		\textbf{d-f}, $P_{ZZ}$ with $S$-like initialisation and \textbf{a}, flipping spin of Q1 adiabatically, \textbf{e}, no spin flip, and \textbf{f}, flipping spin of Q2 adiabatically. All fits are according to Equations~\ref{eq:T1-1}-\ref{eq:T1}. Error range of the $T_1$ numbers represent the 95\% confidence level.
		}
		\label{fig:t1}
\end{extfig*}

\begin{extfig*} 
	\centering 
		\includegraphics[page=9,trim=1.4cm 19cm 1.4cm 0cm]{figures_arxiv.pdf}
		\caption{\textbf{CNOT operation via exchange gate pulsing.}
		\textbf{a}, EDSR spectra of Q1 and Q2 as a function of applied J-gate voltage $\Delta{}V_\text{J}$. At large $\Delta{}V_\text{J}$, the resonance lines clearly split, demonstrating control over the J-coupling.
		\textbf{b}, Pulse sequence of a CNOT-like two-qubit gate.
		\textbf{c},\textbf{d} Measured and simulated parity readout ($P_{ZZ}$) after applying the pulse sequence in \textbf{b}, as a function of $\Delta f_\text{Q2}$ and $\tau_\text{J}$ for $V_\text{CZ}$ = 30~mV. Here, $V_\text{G1}$ is also pulsed at $20$~\% of $V_\text{CZ}$ to maintain a constant charge detuning. The CZ fidelity is $> 90$~\%, confirmed by observing no significant decay over 4 CZ cycles. The simulated Hamiltonian uses $\sigma_{ZI}$ coefficient of 370~kHz and $\sigma_{ZZ}$ coefficient of 89~kHz. The good agreement with the experimental data validates the performance of the CNOT gate.
		\textbf{e,f}, Repeat of \textbf{c,d} with different $V_\text{CZ}$ = 32~mV. The simulated Hamiltonian has $\sigma_{ZI}$ coefficient of 290~kHz and $\sigma_{ZZ}$ coefficient of 135~kHz. Small charge rearrangement in the device has happened between \textbf{c} and \textbf{e}.
		}
		\label{fig:cnot}
\end{extfig*}

\begin{extfig*}  
	\centering 
		\includegraphics[page=10,trim=1.4cm 22cm 1.4cm 0cm]{figures_arxiv.pdf}
		\caption{\textbf{Effective electron temperature of the isolated QD unit cell.}
		\textbf{a}, Charge occupation probability around the (2,4)-(3,3) charge transition, measured through $I_{\text{SET}}$ with a triangular wave applied to $\Delta{}V_\text{G}$. The solid lines are fits to Fermi's distribution, allowing us to extract the effective electron temperature as a function of mixing chamber temperature.
		\textbf{b}, Extracted effective electron temperatures from \textbf{a}. The effective temperature is calculated using the lever arm extracted from \autoref{fig:magneto}.
		The minimum effective electron temperature is $\sim250$~mK at low mixing chamber temperatures. At higher temperatures the effective electron temperature is equal to the mixing chamber temperature.
		Error bars represent the 95\% confidence level.
		}
		\label{fig:temperature}
\end{extfig*}

\begin{extfig*} 
	\centering 
		\includegraphics[page=11,trim=1.4cm 24cm 1.4cm 0cm]{figures_arxiv.pdf}
		\caption{\textbf{Magnetospectroscopy of the (2,4) and (3,3) charge configurations.}
		The transitions that move with magnetic field are caused by Zeeman splitting, allowing us to extract the lever arm of $V_\text{G1}$ to be 0.2128. Since $\Delta V_\text{Gp-p} =  \Delta V_\text{G1} - \Delta V_\text{G3}$, and the pulse is applied symmetrically to both G1 and G3, we can further extract the lever arm of $V_\text{G3}$ to be $0.2128 \times \frac{36.8\text{mV}-20\text{mV}}{40\text{mV}-20\text{mV}} = 0.1788$.
		}
		\label{fig:magneto}
\end{extfig*}

\begin{extfig*} 
	\centering 
		\includegraphics[page=12,trim=1.4cm 22cm 1.4cm 0cm]{figures_arxiv.pdf}
		\caption{\textbf{Magnetic field dependence of qubit properties.}
		\textbf{a}, Spin relaxation time $T_1$, \textbf{b}, Hahn Echo coherence time $T_2^\text{Hahn}$, and \textbf{c}, Ramsey coherence time $T_2^*$ as a function of external magnetic field $B_0$.
		Error bars represent the 95\% confidence level.
		}
		\label{fig:t1t2}
\end{extfig*}

\begin{extfig*} 
	\centering 
		\includegraphics[page=13,trim=6.1cm 19cm 6.1cm 0cm]{figures_arxiv.pdf}
		\caption{\textbf{Readout visibility of the SET charge sensor.}
		\textbf{a-d}, Histograms of charge sensor current $\Delta I_\text{SET} = \bar{I}_\text{SET}(\text{during Read}) - \bar{I}_\text{SET}(\text{during Reset})$ for \textbf{a}, \autoref{fig:hot}a ($T_{\rm MC}=40$~mK, $B_0=0.1$~T), \textbf{b}, \autoref{fig:hot}b ($T_{\rm MC}=40$~mK, $B_0=1.4$~T), \textbf{c}, \autoref{fig:hot}e ($T_{\rm MC}=1.5$~K, $B_0=0.1$~T), and \textbf{d}, \autoref{fig:hot}f ($T_{\rm MC}=1.5$~K, $B_0=1.4$~T). 
		Histograms in \textbf{a,b} are fitted with a Gaussian model including decay from the even parity state to the odd parity state during the readout period~\cite{medford2013}. The extracted visibilities are 88.1\% and 89.3\%, respectively. Assuming no state decay during readout, the ideal readout visibility, which corresponds to the charge readout visibility, would be $V_{\rm ideal}=99.9$\% for $T_{\rm MC}=40$~mK. The histograms in \textbf{c,d} are fitted to the ideal Gaussian model only, revealing $V_{\rm ideal}=78.5$\% and $V_{\rm ideal}=79.5$\%  for $T_{\rm MC}=1.5$~K. This clearly highlights the limitations of SET charge sensing at elevated temperatures, due to the thermal distribution of electrons in the SET source and drain reservoirs.
        }
		\label{fig:histogram}
\end{extfig*}


\end{document}